\documentclass[amsmath,amssymb,nofootinbib]{revtex4}
\usepackage{graphicx}
\usepackage{dcolumn}
\usepackage{color}
\usepackage{bm}
\usepackage{amsmath,amssymb,graphicx}
\usepackage{epsfig}
\usepackage{enumerate}
\usepackage{array}
\usepackage{multirow}
\begin{document}
\title
{Exceptional points and quantum phase transition\\ in a fermionic extension of the Swanson oscillator}
\author{Akash Sinha\footnote{E-mail: akash26121999@gmail.com}$^1$,  Aritra Ghosh\footnote{E-mail: ag34@iitbbs.ac.in}$^1$, Bijan Bagchi\footnote{E-mail: bbagchi123@gmail.com}$^2$}

\vspace{2mm}

\affiliation{$^{1}$School of Basic Sciences, Indian Institute of Technology Bhubaneswar, Jatni, Khurda, Odisha 752050, India\\
$^{2}$Brainware University,
Barasat, Kolkata, West Bengal 700125, India}

\vskip-2.8cm
\date{\today}
\vskip-0.9cm

%--------------------------------------------------------------------------------------------------------------------------------------

\begin{abstract}
Motivated by the structure of the Swanson oscillator which is a well-known example of a non-Hermitian quantum system consisting of a general representation of a quadratic Hamiltonian, we propose a fermionic extension of such a scheme which incorporates two fermionic oscillators together with bilinear-coupling terms that do not conserve particle number. We determine the eigenvalues and eigenvectors, and expose the appearance of exceptional points where two of the eigenstates coalesce with the corresponding eigenvectors exhibiting self-orthogonality with respect to the bi-orthogonal inner product. The model admits a quantum phase transition -- we discuss the two phases and also demonstrate that the ground-state entanglement entropy exhibits a discontinuous jump indicating the transition between the two phases.
 \end{abstract}

\maketitle

\section{Introduction}
In recent times, the study of non-Hermitian systems in quantum mechanics has evinced a lot of interest \cite{NH,NH0,NH1,NH2,NH3,NH4,NH5,NH6}. Parity-time-symmetric Hamiltonians, where the parity operator $\mathcal{P}$ is defined by the operations $(i, x, p) \rightarrow (i, -x, -p)$ and the time-reversal operator $\mathcal{T}$ by the ones $(i, x, p) \rightarrow (-i, x, -p)$, form a distinct subclass of a wider branch of non-Hermitian
Hamiltonians. Such Hamiltonians have drawn considerable attention because a system featuring unbroken \(\mathcal{PT}\)-symmetry generally preserves the reality of the corresponding bound-state eigenvalues \cite{PT1, PT2, BH, PT3}. The last two decades have witnessed the relevance of \(\mathcal{PT}\)-symmetry in optics \cite{MUS}, including non-Hermitian photonics \cite{phot1,phot2} where balancing gain and loss provides a powerful toolbox towards the exploration of new types of light-matter interaction \cite{WANG}.\\

A remarkable feature associated with many non-Hermitian systems is the unique presence of exceptional points which are singular points in the parameter space at which two or more eigenstates (eigenvalues and eigenvectors) coalesce \cite{EP, EP1, KATO, COR, ORTIZ, RAM}. Such points, including the existence of their higher orders \cite{MAND} are of great interest in the context of optics \cite{EPO1,EPO2,EPO3,EPO4} as well as while going for experimental observations in thermal atomic ensembles \cite{EPAMO}. It is worthwhile noting that a non-Hermitian operator (even with real eigenvalues) admits distinct left and right eigenvectors; at the exceptional point, the coalescing eigenvectors become orthogonal to each other, i.e., they exhibit the so-called self-orthogonality condition in which the inner product between the corresponding left and right eigenvectors becomes zero \cite{EP}. This result has found interesting physical implications such as stopping of light in \(\mathcal{PT}\)-symmetric optical waveguides as reported in \cite{lightstops}.  \\ 

A particularly simple yet interesting example of a non-Hermitian system is the Swanson oscillator \cite{swanson1,swanson2,swanson3,fring}, being described by the Hamiltonian (we take \(\hbar = k_B = 1\))
\begin{equation}\label{HB}
H = \omega a^\dagger a + \alpha (a^\dagger)^2 + \beta a^2,
\end{equation} where \(\omega,\alpha, \beta \in \mathbb{R}\), with \(\omega > 0\) and \(\alpha \neq \beta\); the latter condition ensures that the Hamiltonian is non-Hermitian. The Hamiltonian is pseudo-Hermitian \cite{Mostafazadeh,jones,BBpseudo} (and also \(\mathcal{PT}\)-symmetric), and can support a real and positive spectrum. The remarkable feature of the Swanson model is the existence of the terms \((a^\dagger)^2\) and \(a^2\) which are not `number conserving', respectively leading to the transitions \(|n\rangle \mapsto |n+2\rangle\) and \(|n\rangle \mapsto |n-2\rangle\). Exceptional points arising from a situation involving coupled oscillators where each mode is described by a Swanson-like Hamiltonian have been reported recently in \cite{bagchiEP}. \\

In this paper, we present a formalism that addresses a fermionic extension of the Swanson oscillator [Sec. (\ref{modelsection})]. In particular, we demonstrate the existence of exceptional points in the parameter space describing the system; at such points, two of the eigenstates coalesce with the eigenvectors conforming to the self-orthogonality condition with respect to the so-called bi-orthogonal inner product \cite{biortho}, to be defined later [Sec. (\ref{EPsectiongeneral})]. A comparison with the Hermitian case is presented in Sec. (\ref{Hermitiancomparison}). We demonstrate that the fermionic extension admits a quantum phase transition (except in the Hermitian limit) [Sec. (\ref{quantumphasetransitionsec})] which is also observed from a discontinuous jump in the ground-state entanglement entropy [Sec. (\ref{entropysec2})]. We conclude the paper in Sec. (\ref{ConcludeSec}).

\section{The Fermionic Extension: Hamiltonian and Hilbert Space}\label{modelsection}
Towards this end let us consider a quadratic (oscillator) Hamiltonian, but incorporate additional terms that do not lead to conservation of particle number. Since for fermionic operators, say \(c\) and \(c^\dagger\), the properties \(c^2 = (c^\dagger)^2 = 0\) need to be satisfied, a straightforward generalization of Eq. (\ref{HB}) would be quite unfeasible.  However, one could resort to a situation with two fermionic sets of operators \((c_1,c_1^\dagger)\) and \((c_2,c_2^\dagger)\) defining a Hamiltonian that goes as
\begin{equation}\label{HFs}
H = \omega_1 c_1^\dagger c_1 + \omega_2 c_2^\dagger c_2 + \alpha c_1^\dagger c_2^\dagger + \beta  c_2 c_1, \quad \alpha \neq \beta,
\end{equation} where \(\omega_{1,2},\alpha, \beta \in \mathbb{R}\), with \(\omega_{1,2} > 0\). One has the usual anti-commutation relations, i.e., 
\begin{eqnarray}
    \{c_j,c_k^{\dagger}\}=\delta_{j,k},\quad \{c_j,c_k\}=0=\{c_j^{\dagger},c_k^{\dagger}\}, \quad j,k = 1,2.
    \end{eqnarray}  With Eq. (\ref{HFs}) as the candidate for the fermionic extension of the Swanson oscillator, we now proceed to investigate the associated exceptional points which are basically the fingerprints signifying the character of a non-Hermitian system. It may be pointed out that the indices `$1$' and `$2$' can be looked upon as internal indices (like spin, color, etc.) in which case the above Hamiltonian describes a system with a single site which can accommodate two different types of fermions, labeled by `$1$' and `$2$'. Arguably, this interpretation bears a closer resemblance to the bosonic Swanson oscillator. However, as far as our analysis is concerned, such interpretations do not play a role; we thus continue to treat the indices `$1$' and `$2$' as different position labels throughout the rest of the work. \\

     For the fermionic system at hand, the complete Hilbert space can be decomposed as
\begin{eqnarray}
    \mathcal{H}=\mathcal{H}_0\oplus\mathcal{H}_1\oplus\mathcal{H}_2,
\end{eqnarray}
where $\mathcal{H}_0$ and $\mathcal{H}_2$ are one-dimensional (each) and are spanned by the vectors $|\Omega\rangle$ and $c_1^{\dagger}c_2^{\dagger}|\Omega\rangle$, respectively; $\mathcal{H}_1$ is two-dimensional and is spanned by $c_1^{\dagger}|\Omega\rangle$ and $c_2^{\dagger}|\Omega\rangle$. Here \(|\Omega\rangle\) is the zero-particle (vacuum) state. Let us relabel the basis vectors as \( |1\rangle := |\Omega\rangle\), \(|2\rangle := c_1^{\dagger}|\Omega\rangle\), \(|3\rangle := c_2^{\dagger}|\Omega\rangle\), and \( |4\rangle := c_1^{\dagger}c_2^{\dagger}|\Omega\rangle\). In this (natural) basis, the Hamiltonian is expressible as a $4\times 4$ matrix which reads (we pick \(\omega_1 = \omega\) and \(\omega_2 = 1 - \omega\), with \(\omega \in (0,1)\))
\begin{eqnarray}
H=\left(
\begin{array}{cccc}
 0 & 0 & 0 & \alpha  \\
 0 & \omega  & 0 & 0 \\
 0 & 0 & (1-\omega)  & 0 \\
 \beta  & 0 & 0 & 1 \\
\end{array}
\right).\end{eqnarray}
It should be pointed out that for a certain range of the parameters, one can find some matrix \(\eta\) such that \(h = \eta^{-1} H \eta\) is Hermitian, i.e., one can construct a Dyson map. Explicitly, for \(\alpha \beta > 0\), we have
\begin{eqnarray}
    \eta= \left(
\begin{array}{cccc}
 1 & 0 & 0 & (\alpha -\sqrt{\alpha  \beta }) \\
 0 & 1 & 0 & 0 \\
 0 & 0 & 1 & 0 \\
 (\sqrt{\alpha  \beta }-\beta)  & 0 & 0 & 1 \\
\end{array}
\right),
\end{eqnarray} giving
\begin{eqnarray}
    h=
    \left(
\begin{array}{cccc}
 0 & 0 & 0 & \sqrt{\alpha  \beta } \\
 0 & \omega  & 0 & 0 \\
 0 & 0 & (1-\omega)  & 0 \\
 \sqrt{\alpha  \beta } & 0 & 0 & 1 \\
\end{array}
\right),
\end{eqnarray} which is Hermitian. 

\section{Eigenstates, parameter space, and exceptional points}\label{EPsectiongeneral}
For ease of demonstration, we go for the choice \(\omega_1 = \omega\) and \(\omega_2 = 1 - \omega\), with \(\omega \in (0,1)\). The four-dimensional problem admits four eigenstates. Two of the right eigenvectors in the \(\{|1\rangle, |2\rangle, |3\rangle, |4\rangle\}\) basis are
\begin{equation}\label{IandII}
|\psi^{\rm I}_R\rangle =  \begin{pmatrix}
        -\frac{\sqrt{4 \alpha  \beta +1}+1}{2 \beta } \\
        0 \\
        0 \\
       1\\
    \end{pmatrix}, 
    \hspace{5mm}
    |\psi^{\rm II}_R\rangle =  \begin{pmatrix}
        -\frac{1-\sqrt{4 \alpha  \beta +1}}{2 \beta } \\
        0 \\
        0 \\
       1\\
    \end{pmatrix},
\end{equation}
with respective eigenvalues \(E^{\rm I, II} = \frac{1}{2} \left(1\mp\sqrt{4 \alpha  \beta +1}\right)\). The other two eigenvectors are
 \begin{equation}
|\psi^{\rm III}_R\rangle =  \begin{pmatrix}
        0 \\
        1 \\
        0 \\
       0\\
    \end{pmatrix}, 
    \hspace{5mm}
    |\psi^{\rm IV}_R\rangle =  \begin{pmatrix}
        0 \\
        0 \\
        1 \\
       0\\
    \end{pmatrix},
\end{equation} with respective eigenvalues \(E^{\rm III,IV} = \omega, 1-\omega\). 

\subsection{Exceptional points}
The states described by \(|\psi^{\rm III,IV}_R\rangle\) are independent of the `non-Hermiticity' parameters \(\alpha\) and \(\beta\), and therefore cannot be made to coalesce by tuning the parameters \(\alpha\) and \(\beta\). On the other hand, it is clear that the states described by \(|\psi^{\rm I,II}_R\rangle\) depend upon \(\alpha\) and \(\beta\) rather strongly. The corresponding left eigenvectors read 
\begin{equation}
\langle \psi^{\rm I}_L| =  \begin{pmatrix}
        -\frac{\sqrt{4 \alpha  \beta +1}+1}{2 \alpha } \\
        0 \\
        0 \\
       1\\
    \end{pmatrix}^{T}, 
    \hspace{5mm}
    \langle \psi^{\rm II}_L| =  \begin{pmatrix}
        -\frac{1-\sqrt{4 \alpha  \beta +1}}{2 \alpha } \\
        0 \\
        0 \\
       1\\
    \end{pmatrix}^{T}.
\end{equation}
At an exceptional point, it is expected that both the eigenvalues and the eigenvectors coalesce. For the present case, it is found to happen for \(1 + 4 \alpha \beta = 0\), for which \(E^{\rm I} = E^{\rm II} = \frac{1}{2} \) and \(|\psi^{\rm I}_R\rangle=|\psi^{\rm II}_R\rangle\); quite naturally, one also has \(\langle \psi^{\rm I}_L| = \langle \psi^{\rm II}_L|\) which gives (see \cite{biortho} for discussion on the bi-orthogonal inner product which we employ below)
\begin{equation}
\langle \psi^{\rm I,II}_L| \psi^{\rm I,II}_R \rangle =  \frac{1}{4 \alpha \beta} + 1 = 0,
\end{equation}  confirming the self-orthogonality condition \cite{EP}. On the \(\alpha\beta\)-parameter space, the rectangular hyperbola \(4 \alpha \beta + 1=0\) describes the set of points (infinitely many) for which the eigenvalues and eigenvectors coalesce. Thus, the condition \(4 \alpha \beta + 1=0\) may be interpreted as pointing to the `exceptional curve'.

\subsection{\(\alpha\beta\)-parameter space}\label{parametersec}
\begin{figure}
\begin{center}
\includegraphics[scale=0.9]{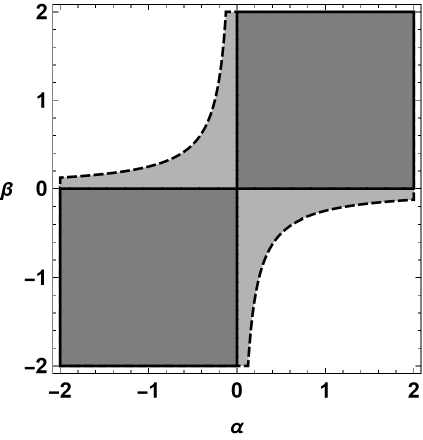}
\caption{Region in the \(\alpha\beta\)-parameter space conforming to \(4\alpha \beta + 1 > 0\) (light and dark gray) and \(\alpha\beta > 0\) (dark gray). The black-dashed curve is \(4\alpha \beta + 1 = 0\).}
\label{ep1234}
\end{center}
\end{figure}

Let us comment on the parameter space which is induced by the parameters \(\alpha\) and \(\beta\) (assuming \(\alpha, \beta \neq 0\)). Since we are looking for real eigenvalues, we shall restrict our attention to the points for which \(4\alpha\beta + 1 \geq 0\). We note that the norm of the states `I' and `II' (in the bi-orthogonal sense) can be determined to be 
\begin{equation}\label{norm111}
\langle \psi^{\rm I}_L| \psi^{\rm I}_R \rangle = 1+ \frac{(1+\sqrt{4 \alpha  \beta +1})^2}{4 \alpha \beta },
\end{equation}
\begin{equation}\label{norm112}
\langle \psi^{\rm II}_L| \psi^{\rm II}_R \rangle = 1 + \frac{(1-\sqrt{4 \alpha  \beta +1})^2}{4 \alpha \beta }.
\end{equation}
Although the norms coalesce and vanish at exceptional points for which \(4\alpha\beta + 1 = 0\), demanding that they are to be positive furnishes the additional condition \(\alpha \beta > 0\). In Fig. (\ref{ep1234}), the region shaded in dark gray (the first and third quadrants excluding the lines \(\alpha = 0\) and \(\beta = 0\)) are where the following two conditions hold: (a) the spectrum is real, (b) the norms are positive. The region shaded in light gray contains those points for which the norms \(\langle \psi^{\rm I}_L| \psi^{\rm I}_R \rangle\) and \(\langle \psi^{\rm II}_L| \psi^{\rm II}_R \rangle\) are not positive-definite, although the spectrum is still real. The exceptional curve \(4\alpha\beta + 1 = 0\) is shown as the dashed curve on which the norms (in the bi-orthogonal sense as defined above) vanish. 

 \subsection{A new set of parameters}\label{Hermitiancomparison}
\begin{figure}[h!]
    \centering
    \includegraphics[scale=0.65]{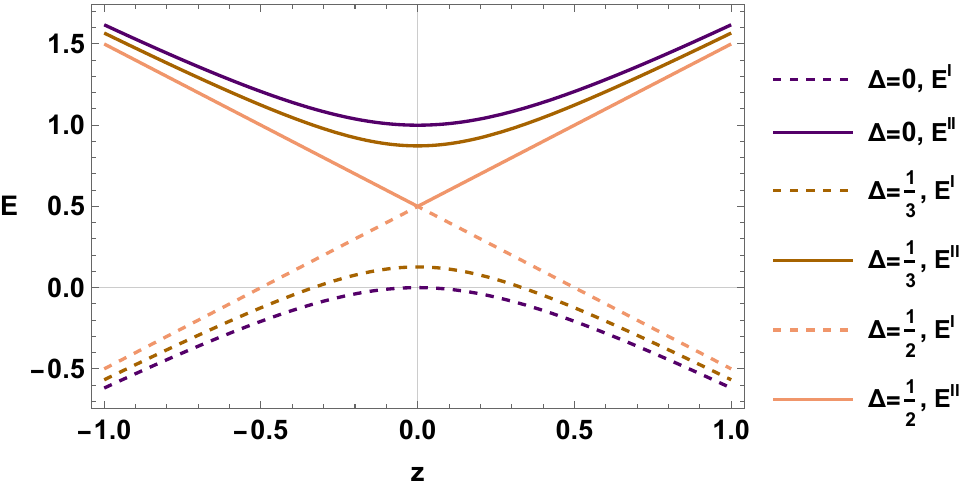}
    \caption{The (real) energy levels $E^{\rm I}$ and $E^{\rm II}$ as functions of $z$ for different values of $\Delta$.} 
    \label{fig:enter-label}
\end{figure}

\begin{figure}[h!]
    \centering
    \includegraphics[scale=0.68]{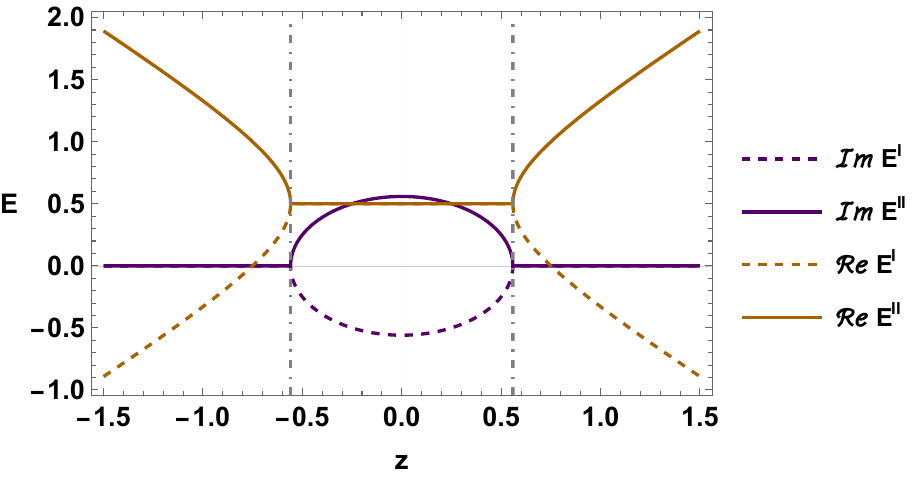}
    \caption{Real and imaginary parts of $E^{\rm I}$ and $E^{\rm II}$ as functions of $z$ for $\Delta = 3/4$.} 
    \label{fig:enter-label1}
\end{figure}

We can introduce a new set of parameters $z$ and $\Delta$ as 
\begin{eqnarray}\label{zdeltadef}
    \alpha=z-\Delta,\quad \beta=z+\Delta,\quad\quad z,\Delta\in{\mathbb R},
\end{eqnarray}
which yields $E^{\rm I,II}=\frac{1}{2}\left(1\mp\sqrt{1+4(z^2-\Delta^2)}\right)$ and the Hermitian limit is achieved by taking $\Delta\to 0$. However, the condition $4\alpha\beta+1\geq 0$ demands that
\begin{eqnarray}
    4(z^2-\Delta^2)\geq-1,
\end{eqnarray}
which translates to 
\begin{eqnarray}
    \Delta^2\leq \frac{1}{4}+z^2.
\end{eqnarray}
Thus, letting \(z\) vary freely implies the following bounds on \(\Delta\):
\begin{eqnarray}
    -\frac{1}{2}\leq \Delta\leq\frac{1}{2}.
\end{eqnarray}
We have plotted $E^{\rm I,II}$ as a function of $z$ in Fig. (\ref{fig:enter-label}) for three different values of $\Delta$. The `exceptional' point is at \(z = 0\) for \(\Delta = 1/2\) at which \(E^{\rm I}\) and \(E^{\rm II}\) meet in Fig. (\ref{fig:enter-label}). It should be pointed out that in Fig. (\ref{fig:enter-label}), we have taken some values of \(\Delta\) such that \(0 \leq \Delta \leq 1/2\) to ensure that the spectrum is real for all \(z\) and for this, we arrive at the exceptional point only at
\(\Delta=1/2\). For larger values of \(\Delta\), the spectrum is not real for all \(z\) but is real only outside a certain domain of \(z\). For instance, we have plotted the real and imaginary parts of \(E^{\rm I}\) and \(E^{\rm II}\) for \(\Delta = 3/4\) in Fig. (\ref{fig:enter-label1}). It is observed that the imaginary parts vanish beyond a certain interval on the \(z\)-axis. The gray-colored vertical lines indicate the exceptional points at which \(E^{\rm I}\) and \(E^{\rm II}\) coalesce for this particular choice of \(\Delta\).

\subsection{Quantum phase transition}\label{quantumphasetransitionsec}
Without any loss of generality, let us take $\omega<\frac{1}{2}$. Some simple manipulations reveal that if $\Delta^2+\omega^2-\omega\geq 0$, there are crossings between the states with energies $\omega$ and $E^{\rm I}=\frac{1}{2}\left(1-\sqrt{1+4(z^2-\Delta^2)}\right)$. The crossovers happen at the points $z_{\pm}=\pm \sqrt{\Delta^2+\omega^2-\omega}$. When $z_-\leq z\leq z_+$, the state with energy $\omega$ becomes the ground state; outside this region, the state having energy $E^{\rm I}$ is the ground state. \\

The ground-state crossings hint towards the possibility of a phase transition in the system \cite{sachdev}. To see this explicitly, we will consider the occupation probabilities of both the sites `1' and `2' for the two different ground states. However, before that one needs to choose an appropriate normalization for the states. It is evident from Eq. (\ref{norm111}) that the norm of the state with energy $E^{\rm I}$ becomes negative for $\alpha\beta<0$. Therefore, the validity of this norm is limited only for the region $\alpha\beta>0$. In terms of $z$ and $\Delta$, this requires
\begin{eqnarray}
    |z|>|\Delta|.
\end{eqnarray}
Unfortunately, we have $|z_{\pm}|<|\Delta|$ if $\omega<1$, and therefore one must resort to a different normalization scheme for further analysis. One possible choice is the standard Dirac norm (see for instance, Ref. \cite{dirac}):
\begin{eqnarray}
    ||\psi||_{L,R}=\sqrt{\langle \psi_{L,R}|\psi_{L,R}\rangle},\quad    \langle{\cal O}\rangle_{L,R}=\langle \psi_{L,R}|{\cal O}|\psi_{L,R}\rangle.
\end{eqnarray}
It is easy to verify that the Dirac norms are positive-semidefinite (zero only if the state itself is trivial); it is for this reason that for subsequent discussions, we shall refer to the Dirac norms and not the bi-orthogonal norms defined in Eqs. (\ref{norm111}) and (\ref{norm112}). \\

One can now distinguish between the two phases by analyzing the occupation probabilities. For $z_-\leq z\leq z_+$, we have 
\begin{eqnarray}
    \langle n_1\rangle_{L,R}=\langle c_1^{\dagger}c_1\rangle_{L,R}=1,\quad \quad\langle n_2\rangle_{L,R}=\langle c_2^{\dagger}c_2\rangle_{L,R}=0,
\end{eqnarray} where the expectation value is taken over the ground state. However, for $|z|>|z_{\pm}|$, we arrive at
\begin{eqnarray}\label{n1n2results}
    \langle n_1\rangle_{L,R}=\langle n_2\rangle_{L,R}=\left[\frac{\left(\sqrt{4 \left(z^2-\Delta^2\right) +1}+1\right)^2}{4 (z\mp\Delta)^2}+1\right]^{-1} ,
\end{eqnarray} wherein, as before the expectation value is taken over the appropriate ground state and the \(\pm\) in Eq. (\ref{n1n2results}) corresponds to `\(L\)' and `\(R\)' Dirac norms, respectively. Therefore, one can consider the first ground state to be localized at the first site while the second ground state is equally spread over both the sites, i.e., it is delocalized. Notice that putting $\Delta=0$ forces $\omega$ to be zero in order to produce real $z_{\pm}$. However, this also yields $z_-=z_+=0$ meaning that the state with energy $\omega$ merely touches the state with energy $E^{\rm I}$ at $z=0$ but one never achieves $\omega< E^{\rm I}$; this can be seen by computing the ground-state occupation probabilities as before for both the regions \(z>0\) and \(z<0\). In other words, the existence of the quantum phase transition is a consequence of the non-Hermiticity of our model. It should be pointed out that the purpose of introducing the parameters \(z\) and \(\Delta\) as defined in Eq. (\ref{zdeltadef}) has been to present a transparent analysis of the spectrum and the two phases, each corresponding to a different ground state; in the next section, we shall revert back to the original parameters \(\alpha\) and \(\beta\).

\section{Ground-state entanglement entropy from Dirac normalization}\label{entropysec2}
The ground state can equivalently be described by the global density matrix that is defined as
\begin{eqnarray}
    \varrho_{L,R}=|G_{L,R}^N\rangle\langle G_{L,R}^N|,
\end{eqnarray} where the superscript `\(N\)' indicates that the state vectors are normalized with respect to the left(\(L\))/right(\(R\)) Dirac norm. To compute the entanglement entropy of, say, the first fermion (index `1'), we need to find the reduced density matrix for the first particle (let us call it $\rho_{L,R}^1$) from which we can compute the von Neumann entropy as
\begin{eqnarray}\label{eq: Neumann entropy}
    S_{L,R}^1=-{\rm Tr}_1[\rho_{L,R}^1 {\rm ln}\rho_{L,R}^1].
\end{eqnarray}
 The usual way to obtain the reduced density matrix describing a particular subsystem demands performing a partial trace on the global density matrix over the rest of the system. Although this is a sensible operation to perform on a system with non-identical particles, the case of identical and indistinguishable particles requires more careful treatment; this is because when one works with a system of indistinguishable particles, the Hilbert space usually has a richer structure as compared to that of a system with non-identical constituents. For example, a system of two non-identical particles is described by the Hilbert space ${\cal H}^1\otimes{\cal H}^2$, where `$1$' and `$2$' label the two particles. However, if the particles obey a specific statistics like Bose or Fermi statistics, the corresponding Hilbert space is not the whole of ${\cal H}^1\otimes{\cal H}^2$, but rather the symmetric and antisymmetric subspaces of ${\cal H}^1\otimes{\cal H}^2$, respectively. This induces intrinsic correlations between different subsystems which arise purely due to the indistinguishability and the statistics of the particles. The Hilbert space for a fermionic (identical) pair is thus defined to be ${\cal H}^1\wedge{\cal H}^2$, where `$\wedge$' denotes antisymmetrization. \\
 
 The traditional approach towards finding the reduced density matrix $\rho_{L,R}^1$ relies rather heavily on the notion of separability of the concerned state. It is usually found by taking the partial trace of the quantity $\varrho_{L,R}$ over ${\cal H}^2$.
Although here we work with two fermionic degrees of freedom, the
separability of the ground state itself is not so transparent; one of the ground states is a linear
superposition of the vacuum state and a two-particle state. This
renders a conceptual difficulty for the analysis of entanglement. To be explicit, the state
\begin{eqnarray}
    |G_R\rangle= \left( -\frac{\sqrt{4 \alpha  \beta +1}+1}{2 \beta }+c_1^{\dagger}c_2^{\dagger}\right)|\Omega\rangle
\end{eqnarray}
 cannot be treated as an element from ${\cal H}^1\wedge{\cal H}^2$  and thus, the idea of performing a partial trace becomes somewhat obscure. In what follows, we shall pursue a different approach to find the reduced density matrix. We will explicitly work with $\varrho_R=|G_R^N\rangle\langle G_R^N|$.

\subsection{Reduced density matrix for fermion `1'} 
The strategy towards finding the reduced density matrix of one of the fermions, say, fermion `1' is as follows (see \cite{ent,rhoref1} for some related discussions). The local Hilbert space is spanned by $|1\rangle:=|\Omega\rangle$ and $|2\rangle:=c_1^{\dagger}|\Omega\rangle$. Thus, any local operator is expressible as 
\begin{eqnarray}
    \mathcal{O}_{a_0,a_1,a_2,a_3} =a_0 \mathbb{I} +a_1 c_1 +a_2 c_1^\dagger + a_3 c_1^\dagger c_1. 
\end{eqnarray}
However, the parity-superselection rule \cite{pss} insists that the (reduced) density matrix must have the following form \cite{bal1,bal2}:
\begin{eqnarray}
     \rho_R^1=a \mathbb{I} +b c_1^\dagger c_1,
     \end{eqnarray}
 for some coefficients \(\{a,b\}\). Now, for a generic local observable \(\mathcal{O}\), one must have
\begin{eqnarray}
    {\rm Tr}_1[\rho_R^1 \mathcal{O}]= \langle G^N_R|\mathcal{O}|G^N_R\rangle = \frac{\langle G_R|\mathcal{O}|G_R\rangle}{\langle G_R|G_R\rangle},
\end{eqnarray} where \({\rm Tr}_1[\cdot]\) is evaluated on the basis \(|\Omega\rangle\) and \(c_1^\dagger|\Omega\rangle\). This serves as a consistency condition allowing one to determine the constants \(\{a,b\}\). \\

\begin{figure}
\begin{center}
\includegraphics[scale=0.67]{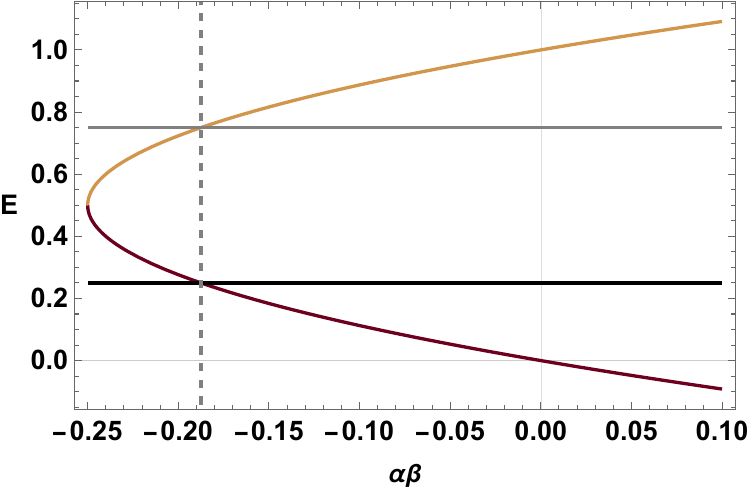}
\caption{Energy eigenvalues \(E^{\rm I}\) (brown), \(E^{\rm II}\) (orange), \(E^{\rm III}\) (black), and \(E^{\rm IV}\) (gray) as a function of \(\alpha \beta\). We have chosen \(\omega = 1/4\). The vertical gray-dashed line corresponds to $\alpha\beta=\omega^2-\omega=-\frac{3}{16}$ which separates the two phases.}
\label{alphabetaspectrum}
\end{center}
\end{figure}

\begin{figure}
\begin{center}
\includegraphics[scale=0.5]{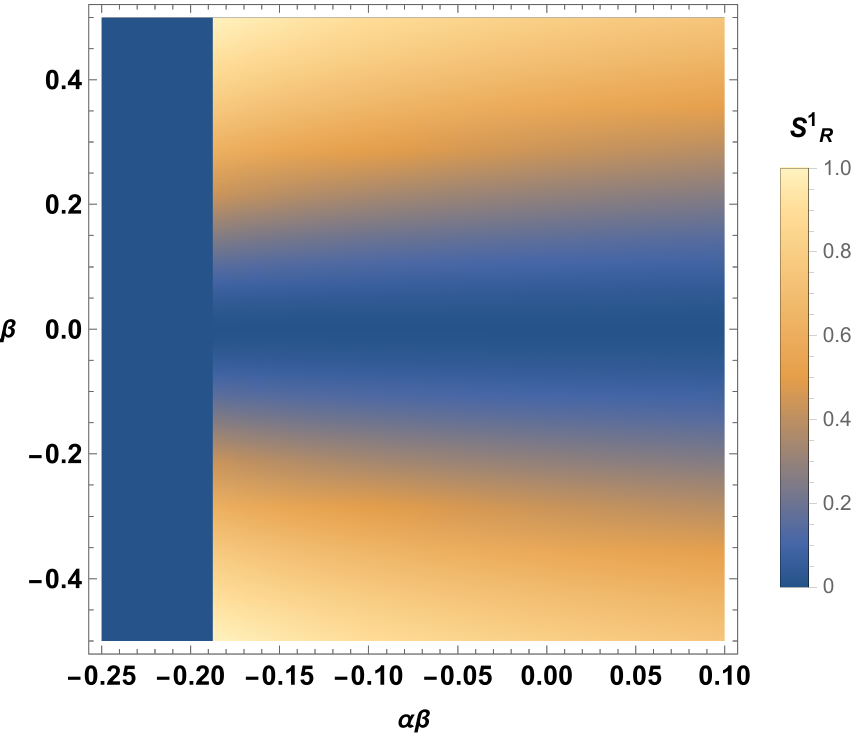}
\caption{Ground-state entanglement entropy. The discontinuous jump as a function of \(\alpha\beta\) is clear.}
\label{entropy111}
\end{center}
\end{figure}

Let us analyze the special case for which $\omega < \frac{1}{2}$; the ground-state energy reads
\begin{eqnarray}
E_G &=& \omega, \hspace{32mm} -\frac{1}{4}<\alpha\beta<\omega^2-\omega, \nonumber \\
&=& \frac{1}{2} \left(1-\sqrt{4 \alpha  \beta +1}\right), \hspace{5mm} \alpha\beta>\omega^2-\omega. 
\end{eqnarray}
 For the purpose of illustration, we have plotted all the four eigenvalues as a function of \(\alpha \beta\) in Fig. (\ref{alphabetaspectrum}) for the choice \(\omega = 1/4\) in which one can observe the ground-state crossing. The ground state reads as
\begin{equation}
|G_R\rangle =  \begin{pmatrix}
        0 \\
        1 \\
        0 \\
       0\\
    \end{pmatrix},  
    \end{equation} for \(-\frac{1}{4}<\alpha\beta<\omega^2-\omega\), and
   \begin{equation}
    | G_R\rangle  =  \begin{pmatrix}
        -\frac{\sqrt{4 \alpha  \beta +1}+1}{2 \beta } \\
        0 \\
        0 \\
       1\\
    \end{pmatrix}, 
\end{equation} for \(\alpha\beta>\omega^2-\omega\). It then simply follows that 
\begin{equation}
    \rho_R^1=
    \begin{pmatrix}
        0 & 0\\
        0 & 1\\
    \end{pmatrix},
    \end{equation}
for \(-\frac{1}{4}<\alpha\beta<\omega^2-\omega\), while 
   \begin{eqnarray}\label{rho1111defffff}
    \rho_R^1=
    \begin{pmatrix}
        \frac{\chi}{1+\chi} & 0\\
        0 & \frac{1}{1+\chi}
    \end{pmatrix}, 
\end{eqnarray} 
where \(\chi = \left(\frac{\sqrt{4 \alpha  \beta +1}+1}{2 \beta }\right)^2\). Notice that \({\rm Tr}_1 [\rho_R^1] = 1\), as anticipated. We have plotted the ground-state entanglement entropy $S_R^1$ as defined by Eq. (\ref{eq: Neumann entropy}) in Fig. (\ref{entropy111}) which shows a discontinuous jump between the two regimes $-\frac{1}{4}<\alpha\beta<\omega^2-\omega$ and $\alpha\beta>\omega^2-\omega$. This confirms the quantum phase transition, being characterized by the discontinuous jump in the entanglement entropy due to the ground-state crossing \cite{sachdev}.\\

It should be emphasized that we have normalized the ground state here with respect to the `right' Dirac norm \(\langle G_R|G_R \rangle\) such that the reduced density matrix so obtained turns out to be positive-semidefinite. One could have alternatively employed the `left' Dirac norm \(\langle G_L|G_L \rangle\) and then, the result for the reduced density matrix and ground-state entanglement entropy would have the same form as obtained above under the interchange \(\alpha \leftrightarrow \beta\). The same phase transition can be observed for both the cases.

\section{Conclusions}\label{ConcludeSec}
We have proposed a fermionic extension of the Swanson oscillator which admits a quadratic but non-Hermitian Hamiltonian by including terms which do not conserve particle number. We have shown that our proposed model admits an infinite number of exceptional points, being given by the points residing on the exceptional curve \(4 \alpha \beta + 1 = 0\) in the parameter space. The model admits a quantum phase transition due to the crossing of the ground states as manifest in the discontinuous jump of the ground-state entanglement entropy. We have shown that the two phases are such that in one of them, the ground state is localized at the first site (index `1') whereas in the other phase, the ground state is equally spread over both the sites. It may be emphasized once again that the phase transition disappears in the Hermitian limit of the model, i.e., the existence of the two distinct phases is a feature due to the non-Hermitian nature of the interaction terms in the Hamiltonian. \\

 Let us end by mentioning that the analysis of this simple two-body problem can also find application
in quantum many-body simulations. For example, let us consider
the following many-body Hamiltonian:
\begin{eqnarray}
    \Tilde{H}=\sum_j\omega_jc_j^{\dagger}c_j+\alpha_jc_j^{\dagger}c_{j+1}^{\dagger}+\beta_jc_{j+1}c_j=\sum_j
h_j.
\end{eqnarray}
It is now well known that the time-evolution operator $U(t)=\exp[-i\Tilde{H}t]$ can be approximated by Trotterization \cite{comp0,comp01}, i.e., by approximating the actual evolution operator by the following expression:
\begin{eqnarray}
  U(t) \approx  \left(\prod_j \exp[-ih_j t/N]\right)^N,
\end{eqnarray} which becomes exact in the limit $N\to\infty$. This
particular approach has found its usage in several contexts \cite{comp1,comp2,comp3}. Notably, the $h_j$'s in the above expression are essentially similar to the two-fermion model discussed in this paper and thus, understanding the physics of our simple two-body system may help in the analysis of more complicated systems as motivated above.\\

%\textbf{Author contributions:} All authors listed have made a substantial, direct, and intellectual contribution to the work and approved it for publication.\\

%\textbf{Data availability statement:} We do not analyze or generate any datasets, because our work proceeds within a theoretical and mathematical approach.\\

\textbf{Acknowledgements:} 
We thank Jasleen Kaur for carefully reading the manuscript. A.S. thanks P. Padhmanabhan and A.P. Balachandran for useful discussions, and also acknowledges the financial support from IIT Bhubaneswar in the form of an Institute Research Fellowship. The work of A.G. is supported by Ministry of Education (MoE), Government of India in the form of a Prime Minister's Research Fellowship (ID: 1200454). B.B. thanks Brainware University for infrastructural support. %We are grateful to the anonymous referees for their valuable inputs which have led to a substantial improvement of the paper. 

%\textbf{Conflict of interest:} The authors declare that the research was conducted in the absence of any commercial or financial relationships that could be construed as a potential conflict of interest.


\begin{thebibliography}{99}
%%%%%%%%%%%%%%%%%%%%%%%%%%%%%%%%

  
   \bibitem{NH}
    I. Rotter,
   	J. Phys. A: Math. Theor. \textbf{42}, 153001 (2009).
  
   \bibitem{NH0}
    Y. Michishita and R. Peters,
   	Phys. Rev. Lett. \textbf{124}, 196401 (2020).
  
  
    \bibitem{NH1}
    K. Holmes, W. Rehman, S. Malzard, and E.-M. Graefe,
   	Phys. Rev. Lett. \textbf{130}, 157202 (2023).




 \bibitem{NH2}
    E.-M. Graefe, M. H\"oning, and H. J. Korsch,
   	J. Phys. A: Math. Theor. \textbf{43}, 075306 (2010).
	
	 \bibitem{NH3}
    \'A. G\'omez-Le\'on, T. Ramos, A. Gonz\'alez-Tudela, and D. Porras,
   	Phys. Rev. A \textbf{106}, L011501 (2022).


 \bibitem{NH4}
    X. Niu, J. Li, S. L. Wu, and X. X. Yi,
   	Phys. Rev. A \textbf{108}, 032214 (2023).
	
	 \bibitem{NH5}
    S. Yao and Z. Wang,
   	Phys. Rev. Lett. \textbf{121}, 086803 (2018).
	
	 \bibitem{NH6}
    N. Okuma and M. Sato,
   	Annu. Rev. Condens. Matter Phys. \textbf{14}, 83 (2023). 

	
	
	
	
	
	
	 \bibitem{PT1}
    C. M. Bender and S. Boettcher,
   	Phys. Rev. Lett. \textbf{80}, 5243 (1998).
	
	 \bibitem{PT2}
    C. M. Bender, S. Boettcher, and P. N. Meisinger,
   	J. Math. Phys. \textbf{40}, 2201 (1999).

    \bibitem{BH}
    C. M. Bender and D. W. Hook,
   	arXiv: 2312.17386.


 \bibitem{PT3}
    R. El-Ganainy, K. G. Makris, M. Khajavikhan, Z. H. Musslimani, S. Rotter, and D. N. Christodoulides,
   	Nat. Phys. \textbf{14}, 11 (2018).
    
 \bibitem{MUS}
    Z. H. Musslimani, K. G. Makris, R. El-Ganainy, and D. N. Christodoulides, 
Phys. Rev. Lett. \textbf{100}, 030402 (2008).

 \bibitem{phot1}
  L.  Feng, R. El-Ganainy, and L. Ge,
   	Nat. Photonics \textbf{11}, 752 (2017).


 \bibitem{phot2}
  R. El-Ganainy, M. Khajavikhan, D. N. Christodoulides, and S. K. Ozdemir,
   	Commun. Phys. \textbf{2}, 37 (2019).
	
	

     \bibitem{WANG}
     C. Wang, Z. Fu, W. Mao, J. Qie, A. D. Stone, and L. Yang,
     Adv. Opt. Photonics \textbf{15}, 442 (2023).
     

 \bibitem{EP}
  W. D. Heiss,
   	J. Phys. A: Math. Theor. \textbf{45}, 444016 (2012).

    
 \bibitem{EP1}
  M. Znojil,
        J. Math. Phys. \textbf{62}, 052103 (2021).

        \bibitem{KATO}
 T. Kato, 
 {\it Perturbation Theory for Linear Operators: Classics in Mathematics}, 
 Springer (1995).

    \bibitem{COR}
 F. Correa  and M. S. Plyushchay, 
 Phys. Rev. D \textbf{86}, 085028 (2012).

 \bibitem{ORTIZ} 
    K. Zelaya and O. Rosas-Ortiz,
    Quantum Rep. \textbf{3}, 458 (2021).

\bibitem{RAM} 
V. Fern\'{a}ndez, R. Ram\'irez, and M. Reboiro, 
J. Phys. A: Math. Theor. \textbf{55}, 015303 (2022).

\bibitem{MAND} 
I. Mandal and E. J. Bergholtz, 
Phys. Rev. Lett. \textbf{127}, 186601 (2021).

 \bibitem{EPO1}
  M.-A. Miri and A. Al\`u,
   	Science \textbf{363}, eaar7709 (2019). 

 \bibitem{EPO2}
  Y. Zhiyenbayev, Y. Kominis, C. Valagiannopoulos, V. Kovanis, and A. Bountis,
   	Phys. Rev. A \textbf{100}, 043834 (2019).
	
 \bibitem{EPO3}
  J. Wiersig,
   	Photonics Res. \textbf{8}, 1457 (2020).
	
	
 \bibitem{EPO4}
  A. Li, H. Wei, M. Cotrufo, W. Chen, S. Mann, X. Ni, B. Xu, J. Chen, J. Wang, S. Fan, C.-W. Qiu, A. Al\`u, and L. Chen,
   	Nat. Nanotechnol. \textbf{18}, 706 (2023).
	
	
	

	
	 \bibitem{EPAMO}
 C. Liang, Y. Tang, A.-N. Xu, and Y.-C. Liu,
   	Phys. Rev. Lett. \textbf{130}, 263601 (2023).

\bibitem{lightstops}
 T. Goldzak, A. A. Mailybaev, and N. Moiseyev,
   	Phys. Rev. Lett. \textbf{120}, 013901 (2018).
	
	 \bibitem{swanson1}
    M. S. Swanson,
   	J. Math. Phys. \textbf{45}, 585 (2004).

  
	 \bibitem{swanson2}
    E.-M. Graefe, H. J. Korsch, A. Rush, and R. Schubert,
   	J. Phys. A: Math. Theor. \textbf{48}, 055301 (2015).
	
	 \bibitem{swanson3}
    B. Bagchi and I. Marquette,
   	Phys. Lett. A \textbf{379}, 1584 (2015). 

    \bibitem{fring} A. Fring and M. H. Y. Moussa, 
    Phys. Rev. A \textbf{94}, 042128 (2016).
	
	\bibitem{Mostafazadeh}
    A. Mostafazadeh,
   	J. Math. Phys. \textbf{43}, 205 (2002).
	
	\bibitem{jones}
    H. F. Jones,
   	J. Phys. A: Math. Gen. \textbf{38}, 1741 (2005).
	
	\bibitem{BBpseudo}
    B. Bagchi, C. Quesne, and R. Roychoudhury,
   	J. Phys. A: Math. Gen. \textbf{38}, L647 (2005).
	
	\bibitem{bagchiEP}
   B. Bagchi, R. Ghosh, and S. Sen,
   	EPL \textbf{137}, 50004 (2022).
	
	
	\bibitem{biortho}
  D. C. Brody, 
  J. Phys. A: Math. Theor. \textbf{47}, 035305 (2014).


		\bibitem{sachdev}
  S. Sachdev,
  {\it Quantum Phase Transitions}, 2nd. ed., Cambridge University Press (2011).



	\bibitem{dirac}
  F. Ruzicka, K. S. Agarwal, and Y. N. Joglekar,
   	J. Phys.: Conf. Ser. \textbf{2038}, 012021 (2021).
	
	
	
		\bibitem{ent}
  A. F. Reyes-Lega,
  {\it Some aspects of operator algebras in quantum physics},
   	Geometric, Algebraic, and Topological Methods for Quantum Field Theory: Proceedings of the 2013 Villa de Leyva Summer School, Villa de Leyva, Colombia, 15–27 July 2013, 1-74 (2017).

	
	
	
	\bibitem{rhoref1}
 L. Herviou, N. Regnault, and J. H. Bardarson,
   	SciPost Phys. \textbf{7}, 069 (2019).



 \bibitem{pss}
G. C. Wick, A. S. Wightman, and E. P. Wigner,
 Phys. Rev. \textbf{88}, 101 (1952).




	  \bibitem{bal1}
  A. P. Balachandran, T. R. Govindarajan, A. R. de Queiroz, and A. F. Reyes-Lega,
Phys. Rev. Lett. \textbf{110}, 080503 (2013).

 \bibitem{bal2}
 A. P. Balachandran, T. R. Govindarajan, A. R. de Queiroz, and A. F. Reyes-Lega,
 Phys. Rev. A \textbf{88}, 022301 (2013).



 \bibitem{comp0}
 H. F. Trotter,
Proc. Am. Math. Soc. \textbf{10}, 545 (1959).

\bibitem{comp01}
M. Suzuki,
Phys. Lett. A \textbf{180}, 232 (1993). 





 \bibitem{comp1}
 B. P. Lanyon, C. Hempel, D. Nigg, M. M\"uller, R. Gerritsma,
F. Z\"ahringer, P. Schindler, J. T. Barreiro, M. Rambach, G.
Kirchmair, M. Hennrich, P. Zoller, R. Blatt, and C. F. Roos,  
Science \textbf{334}, 57 (2011). 



 \bibitem{comp2}
 R. Barends, L. Lamata, J. Kelly, L. Garc\'ia-\'Alvarez, A. G. Fowler, A. Megrant, E. Jeffrey, T. C. White, D. Sank, J. Y. Mutus, B. Campbell, Y. Chen, Z. Chen, B. Chiaro, A. Dunsworth, I.-C. Hoi, C. Neill, P. J. J. O’Malley, C. Quintana, P. Roushan, A. Vainsencher, J. Wenner, E. Solano, and J. M. Martinis,
 Nat. Commun. \textbf{6}, 7654 (2015).



 \bibitem{comp3}
M. Heyl, P. Hauke, and P. Zoller,
Sci. Adv. \textbf{5}, eaau8342 (2019).



\end{thebibliography}
\end{document}